\documentclass[dvips]{acta}
\usepackage{supertabular,lscape,epsfig}
\usepackage{amssymb}
\usepackage{amsmath}
\DeclareSymbolFont{ppa}{OT1}{ppl}{m}{it}
\DeclareMathSymbol{\vv}{\mathalpha}{ppa}{'166}

\newfont{\hb}{rphvb at 10pt}
\newfont{\hbo}{rphvbo at 10pt}
\newfont{\bitt}{rptmbi at 12pt}
\newfont{\bits}{rptmbi at 11pt}

\SetPages{131}{147}

\SetVol{66}{2016}

\begin{document}

\newcommand{\TabCapp}[2]{\begin{center}\parbox[t]{#1}{\centerline{
  \small {\spaceskip 2pt plus 1pt minus 1pt T a b l e}
  \refstepcounter{table}\thetable}
  \vskip2mm
  \centerline{\footnotesize #2}}
  \vskip3mm
\end{center}}

\newcommand{\TTabCap}[3]{\begin{center}\parbox[t]{#1}{\centerline{
  \small {\spaceskip 2pt plus 1pt minus 1pt T a b l e}
  \refstepcounter{table}\thetable}
  \vskip2mm
  \centerline{\footnotesize #2}
  \centerline{\footnotesize #3}}
  \vskip1mm
\end{center}}

\newcommand{\MakeTableSepp}[4]{\begin{table}[p]\TabCapp{#2}{#3}
  \begin{center} \TableFont \begin{tabular}{#1} #4
  \end{tabular}\end{center}\end{table}}

\newcommand{\MakeTableee}[4]{\begin{table}[htb]\TabCapp{#2}{#3}
  \begin{center} \TableFont \begin{tabular}{#1} #4
  \end{tabular}\end{center}\end{table}}

\newcommand{\MakeTablee}[5]{\begin{table}[htb]\TTabCap{#2}{#3}{#4}
  \begin{center} \TableFont \begin{tabular}{#1} #5
  \end{tabular}\end{center}\end{table}}

\newfont{\bb}{ptmbi8t at 12pt}
\newfont{\bbb}{cmbxti10}
\newfont{\bbbb}{cmbxti10 at 9pt}
\newcommand{\uprule}{\rule{0pt}{2.5ex}}
\newcommand{\douprule}{\rule[-2ex]{0pt}{4.5ex}}
\newcommand{\dorule}{\rule[-2ex]{0pt}{2ex}}
\def\thefootnote{\fnsymbol{footnote}}
\begin{Titlepage}
\Title{The OGLE Collection of Variable Stars.\\
Over 45\,000 RR~Lyrae Stars in the Magellanic System\footnote{Based on observations
obtained with the 1.3-m Warsaw telescope at the Las Campanas Observatory of
the Carnegie Institution for Science.}}
\Author{I.~~S~o~s~z~y~ñ~s~k~i$^1$,~~
A.~~U~d~a~l~s~k~i$^1$,~~
M.\,K.~~S~z~y~m~a~ñ~s~k~i$^1$,~~
£.~~W~y~r~z~y~k~o~w~s~k~i$^1$,\\
K.~~U~l~a~c~z~y~k$^{1,2}$,~~
R.~~P~o~l~e~s~k~i$^{1,3}$,~~
P.~~P~i~e~t~r~u~k~o~w~i~c~z$^1$,~~
S.~~K~o~z~³~o~w~s~k~i$^1$,\\
D.\,M.~~S~k~o~w~r~o~n$^1$,~~
J.~~S~k~o~w~r~o~n$^1$,~~
P.~~M~r~ó~z$^1$~~
and~~M.~~P~a~w~l~a~k$^1$}
{$^1$Warsaw University Observatory, Al.~Ujazdowskie~4, 00-478~Warszawa, Poland\\
e-mail: (soszynsk,udalski)@astrouw.edu.pl\\
$^2$ Department of Physics, University of Warwick, Gibbet Hill Road, Coventry, CV4~7AL,~UK\\
$^3$ Department of Astronomy, Ohio State University, 140 W. 18th Ave., Columbus, OH~43210,~USA}
\Received{June 21, 2016}
\end{Titlepage}

\Abstract{We present the largest collection of RR~Lyrae stars in the
  Magellanic System and in its foreground. The sample consists of 45\,451
  RR~Lyr stars, of which 39\,082 were detected toward the Large Magellanic
  Cloud and 6369 toward the Small Magellanic Cloud. We provide long-term
  time-series photometric measurements collected during the fourth phase of
  the Optical Gravitational Lensing Experiment (OGLE-IV).

  We discuss several potential astrophysical applications of our
  collection: investigation of the structure of the Magellanic Clouds and
  the Galactic halo, studies of the globular clusters in the Magellanic
  System, analysis of double-mode RR~Lyr stars, and search for RR~Lyr stars
  in eclipsing binary systems.}{Stars: variables: RR~Lyrae -- Stars:
  oscillations -- Stars: Population II -- Magellanic Clouds -- Catalogs}

\vspace*{5mm}
\Section{Introduction}
RR~Lyrae stars are a powerful tool to trace the oldest ($\ge10$~Gyr)
stellar component in our and other galaxies. These radially pulsating stars
are numerous, easily identifiable, and present in all Local Group galaxies,
irrespective of their morphological type. RR~Lyr stars are standard candles
making them important distance indicators, probes of the three-dimensional
structures of their parent galaxies, and tracers of the star formation
history. Large samples of RR~Lyr variables have proven useful in
investigating the metallicity distribution in galaxies and in determination
of the interstellar extinction maps.

The Optical Gravitational Lensing Experiment (OGLE, Udalski, Szymañski and
Szymañski 2015) has already released catalogs of RR~Lyr stars in the Large
(LMC) and Small Magellanic Clouds (SMC) based on the observations obtained
in 1997--2000 during the OGLE-II (Soszyñski \etal 2002, 2003) and in
2001--2009 during the OGLE-III phases (Soszyñski \etal 2009, 2010). The
latter release contained a total of 27\,381 RR~Lyr stars detected in
54~square degrees of the sky covering central regions of the LMC and
SMC. These catalogs found many astrophysical applications. They were used
by various authors to investigate three-dimensional structure of the
Magellanic Clouds (Pejcha and Stanek 2009, Subramaniam and Subramanian
2009, Haschke \etal 2012b, Kapakos and Hatzidimitriou 2012, Subramanian and
Subramaniam 2012, Wagner-Kaiser and Sarajedini 2013, Moretti \etal 2014,
Deb and Singh 2014, Deb \etal 2015), to examine metallicity gradients in
both galaxies (Feast \etal 2010, Kapakos \etal 2011, Kapakos and
Hatzidimitriou 2012, Haschke \etal 2012a, Wagner-Kaiser and Sarajedini
2013), to construct reddening maps toward both Clouds (Pejcha and Stanek
2009, Haschke \etal 2011, Wagner-Kaiser and Sarajedini 2013), to study the
RR~Lyr light curve morphology in various passbands (Chen \etal 2013,
Gavrilchenko \etal 2014, Moretti \etal 2014), and to analyze relations
between periods, luminosities, colors, amplitudes, and metallicities of
RR~Lyr stars (Ripepi \etal 2012, Bhardwaj \etal 2014, Muraveva \etal
2015). The OGLE samples were also used as training sets for the automatic
systems of the variable star classification (Long \etal 2012, Kim \etal
2014, Kim and Bailer-Jones 2016).

In this paper, we present a new OGLE collection of RR~Lyr stars in the
Magellanic System being an extension of the OGLE-III catalog to the regions
covered by the OGLE-IV fields. About 650 square degrees of the sky
regularly monitored by the OGLE-IV survey cover a large part of the
Magellanic System, including the outskirts of the two galaxies and the
Magellanic Bridge connecting them. The new OGLE release of RR~Lyr variables
in the Magellanic System contains 45\,451 objects in total. This is the
largest set of RR~Lyr stars published to date in any stellar environment.

The paper is organized as follows. Section~2 presents observational data
used in this investigation. Methods used in the selection and
classification of RR~Lyr stars are detailed in Section~3. The collection
itself is described in Section~4. In Section~5, we estimate the
completeness of our sample and compare it with other catalogs of RR~Lyr
stars in the Magellanic Clouds. In Section~6, we discuss some possible
applications of our collection. Finally, conclusions are presented in
Section~7.
\vspace*{-3pt}
\Section{Observations and Data Reduction}
\vspace*{-6pt}
The time-series photometry used in this study has been obtained with the
1.3-m Warsaw telescope at Las Campanas Observatory (operated by the
Carnegie Institution for Science), Chile, between March 2010 and July
2015. The telescope is equipped with a mosaic camera composed of 32 CCDs,
each with 2048 by 4096 pixels, providing a field of view of 1.4~square
degrees on the sky. Most of the observations were obtained through the
Cousins {\it I} filter -- typically from 100 to 750 points, depending on
the field. In the Johnson {\it V}-band we secured from several to 260
observations for color information.

Altogether 475 OGLE-IV fields cover about 650 square degrees in the
Magellanic System, including the Magellanic Bridge between the LMC and SMC
and selected peripheral areas, up to 20~degrees from the centers of the
galaxies. The total number of point sources in the Magellanic Cloud OGLE-IV
database exceeds 75 million. The OGLE data reduction pipeline is based on
the Difference Image Analysis technique (Alard and Lupton 1998, Wo¼niak
2000). The reduction procedures, photometric calibrations and astrometric
transformations have been described by Udalski \etal (2015a).
\vspace*{-3pt}
\Section{Selection and Classification of RR~Lyrae Stars}
\vspace*{-6pt}
We performed a period search for nearly all {\it I}-band light curves
stored in the OGLE database. The only cut was done on the number of data
points that had to be larger than 30. We used the {\sc Fnpeaks}
code\footnote{\it
  http://helas.astro.uni.wroc.pl/deliverables.php?lang=en\&active=fnpeaks}
which calculates the Fourier amplitude spectra and provides the best
periods with their signal-to-noise ratios. For each star we derived two
periods, the second one on the residual light curve obtained by the
subtraction of the primary periodicity.

A search for RR~Lyr stars was conducted on light curves with periods
between 0.2 and 1~day. The preselection of the candidates was based on the
Fourier decomposition of the light curves (Simon and Lee 1981) and the
template fitting to the {\it I}-band light curves. Double-mode RR~Lyr stars
were identified on the basis of their period ratios, which fall in a narrow
range of 0.72--0.75. However, the automatic algorithms played only a
supporting role in the process of the manual selection and classification
of RR~Lyr stars. The final decision on each object was made after a visual
inspection of its light curve. In doubtful cases we took into account the
position of the star in the color--magnitude diagram, as well as
period--luminosity, period--amplitude, and other diagrams.

The selected sample of RR~Lyr variables has been divided into three
classes: fundamental-mode RRab stars, first-overtone RRc stars, and
double-mode RRd stars. Our classification was based on the periods,
amplitudes, and light curve shapes of the stars. In several dozen cases we
corrected the pulsation modes provided by Soszyñski \etal (2009, 2010). All
objects classified in the OGLE-III catalogs as RRe stars (second-overtone
pulsators) have been incorporated to the RRc group. The second-overtone
RR~Lyr stars have not been separated from the first-overtone variables
because of the doubts whether the RRe stars exist at all. We did not find
any natural boundary, which could be used for an unambiguous separation of
the first- and second-overtone RR~Lyr stars.

\MakeTableSep{
l@{\hspace{8pt}} c@{\hspace{6pt}} | l@{\hspace{8pt}}
c@{\hspace{6pt}}}{12.5cm} {Reclassified variables from the OGLE-III
catalogs of RR~Lyr stars in the Magellanic Clouds} 
{\hline
\noalign{\vskip3pt}
\multicolumn{1}{c}{Identifier} & New            & \multicolumn{1}{c}{Identifier} & New   \\
                               & classification &                                & classification \\
\noalign{\vskip3pt}
\hline
\noalign{\vskip3pt}
OGLE-LMC-RRLYR-00077 & Other          & OGLE-LMC-RRLYR-13259 & Eclipsing  \\
OGLE-LMC-RRLYR-00485 & Other          & OGLE-LMC-RRLYR-13512 & Eclipsing  \\
OGLE-LMC-RRLYR-00803 & Spotted        & OGLE-LMC-RRLYR-15806 & Other      \\
OGLE-LMC-RRLYR-00824 & Other          & OGLE-LMC-RRLYR-16124 & Eclipsing  \\
OGLE-LMC-RRLYR-00961 & Eclipsing      & OGLE-LMC-RRLYR-16210 & Eclipsing  \\
OGLE-LMC-RRLYR-01104 & Eclipsing      & OGLE-LMC-RRLYR-16426 & Other      \\
OGLE-LMC-RRLYR-01257 & Other          & OGLE-LMC-RRLYR-16656 & Eclipsing  \\
OGLE-LMC-RRLYR-01802 & Eclipsing      & OGLE-LMC-RRLYR-16795 & Eclipsing  \\
OGLE-LMC-RRLYR-02171 & Other          & OGLE-LMC-RRLYR-17073 & Eclipsing  \\
OGLE-LMC-RRLYR-02376 & Other          & OGLE-LMC-RRLYR-17117 & Eclipsing  \\
OGLE-LMC-RRLYR-02390 & Eclipsing      & OGLE-LMC-RRLYR-17396 & Eclipsing  \\
OGLE-LMC-RRLYR-02548 & Eclipsing      & OGLE-LMC-RRLYR-17584 & Eclipsing  \\
OGLE-LMC-RRLYR-03158 & Eclipsing      & OGLE-LMC-RRLYR-17847 & Other      \\
OGLE-LMC-RRLYR-03802 & Eclipsing      & OGLE-LMC-RRLYR-18086 & Other      \\
OGLE-LMC-RRLYR-04103 & Eclipsing      & OGLE-LMC-RRLYR-18360 & Other      \\
OGLE-LMC-RRLYR-04426 & Classical Cep. & OGLE-LMC-RRLYR-18854 & Other      \\
OGLE-LMC-RRLYR-04733 & Other          & OGLE-LMC-RRLYR-19067 & Eclipsing  \\
OGLE-LMC-RRLYR-04862 & Eclipsing      & OGLE-LMC-RRLYR-19207 & Other      \\
OGLE-LMC-RRLYR-04892 & Eclipsing      & OGLE-LMC-RRLYR-19243 & Eclipsing  \\
OGLE-LMC-RRLYR-05128 & Eclipsing      & OGLE-LMC-RRLYR-19258 & Other      \\
OGLE-LMC-RRLYR-05282 & Other          & OGLE-LMC-RRLYR-19438 & Eclipsing  \\
OGLE-LMC-RRLYR-05305 & Eclipsing      & OGLE-LMC-RRLYR-20089 & Other      \\
OGLE-LMC-RRLYR-05784 & Eclipsing      & OGLE-LMC-RRLYR-20767 & Eclipsing  \\
OGLE-LMC-RRLYR-06232 & Eclipsing      & OGLE-LMC-RRLYR-20781 & Other      \\
OGLE-LMC-RRLYR-06645 & Other          & OGLE-LMC-RRLYR-20821 & Eclipsing  \\
OGLE-LMC-RRLYR-07073 & Other          & OGLE-LMC-RRLYR-21161 & Eclipsing  \\
OGLE-LMC-RRLYR-07569 & Eclipsing      & OGLE-LMC-RRLYR-21207 & Other      \\
OGLE-LMC-RRLYR-07905 & Eclipsing      & OGLE-LMC-RRLYR-21255 & Eclipsing  \\
OGLE-LMC-RRLYR-07935 & Other          & OGLE-LMC-RRLYR-21285 & Eclipsing  \\
OGLE-LMC-RRLYR-08281 & Other          & OGLE-LMC-RRLYR-21455 & Eclipsing  \\
OGLE-LMC-RRLYR-08457 & Other          & OGLE-LMC-RRLYR-22035 & Other      \\
OGLE-LMC-RRLYR-09009 & Eclipsing      & OGLE-LMC-RRLYR-22482 & Eclipsing  \\
OGLE-LMC-RRLYR-09044 & Eclipsing      & OGLE-LMC-RRLYR-22492 & Other      \\
OGLE-LMC-RRLYR-09614 & Other          & OGLE-LMC-RRLYR-23055 & Other      \\
OGLE-LMC-RRLYR-10747 & Other          & OGLE-LMC-RRLYR-23485 & Other      \\
OGLE-LMC-RRLYR-10933 & Eclipsing      & OGLE-LMC-RRLYR-23513 & Eclipsing  \\
OGLE-LMC-RRLYR-10994 & Eclipsing      & OGLE-LMC-RRLYR-23517 & Eclipsing  \\
OGLE-LMC-RRLYR-11143 & Other          & OGLE-LMC-RRLYR-23685 & Other      \\
OGLE-LMC-RRLYR-11169 & Eclipsing      & OGLE-LMC-RRLYR-23868 & Other      \\
OGLE-LMC-RRLYR-11606 & Other          & OGLE-LMC-RRLYR-24247 & Other      \\
OGLE-LMC-RRLYR-12072 & Eclipsing      & OGLE-LMC-RRLYR-24338 & Other      \\
OGLE-LMC-RRLYR-12151 & Eclipsing      & OGLE-LMC-RRLYR-24428 & Eclipsing  \\
OGLE-LMC-RRLYR-12343 & Other          & OGLE-SMC-RRLYR-0251  & Other      \\
OGLE-LMC-RRLYR-12875 & Eclipsing      & OGLE-SMC-RRLYR-0720  & Other      \\
\noalign{\vskip3pt}
\hline}

As a result, we found 18\,158 RR~Lyr stars in both Clouds that were not
recorded during previous stages of the OGLE survey. Only 227 of them (less
than 1\% of the OGLE-III sample) were found in the region covered by the
OGLE-III fields, confirming the high completeness of the OGLE collection of
variable stars. Other newly detected variables lie in the outer regions of
the two galaxies. We also verified the OGLE-III samples of RR~Lyr stars
(Soszyñski \etal 2009, 2010) with the OGLE-IV photometry and we decided to
remove 88 of the sources previously classified as RR~Lyr stars (0.3\% of
the original sample). The detailed list of these objects is given in
Table~1. Most of them turned out to be eclipsing or ellipsoidal binary
systems or their variability type could not be unambiguously identified from
the OGLE photometry (they are designated in Table~1 as ``Other''). Another
$\approx200$ stars in our collection should be treated with caution, because
their classification is uncertain. These stars are flagged in appropriate
data files of our collection.

\Section{RR~Lyrae Stars in the Magellanic System}
The current version of the OGLE collection of RR~Lyr stars in the
Magellanic System contains variables detected during the previous stages of
the OGLE project (Soszyñski \etal 2002, 2003, 2009, 2010) and during the
current, fourth phase of the survey. The whole sample consists of 45\,451
RR~Lyr stars (32\,581 RRab, 10\,246 RRc, and 2624 RRd stars, including 22
anomalous RRd stars, Soszyñski \etal 2016, in preparation), of which 39\,082 and
6369 variables were found toward the LMC and SMC, respectively. The
on-sky boundary between both Clouds may be established only approximately,
because the halos of the two galaxies overlap with each other. In our
collection, we adopted celestial meridian of 2\zdot\uph8 to separate the
LMC and SMC samples, because we found a local minimum in the number of
RR~Lyr variables around this value of right ascension. Our collection also
includes RR~Lyr stars from the halo of the Milky Way. It is impossible to
unambiguously separate Galactic and Magellanic Cloud old stellar
populations, since the Clouds are immersed in the halo of our Galaxy. There
is no a natural luminosity boundary that separates RR~Lyr stars from the
Magellanic Clouds and the Milky Way.

The entire collection can be downloaded from the OGLE Internet Archive
through anonymous FTP sites or via a web interface:
\begin{center}
{\it ftp://ftp.astrouw.edu.pl/ogle/ogle4/OCVS/lmc/rrlyr/}\\ 
{\it ftp://ftp.astrouw.edu.pl/ogle/ogle4/OCVS/smc/rrlyr/}\\ 
{\it http://ogle.astrouw.edu.pl}
\end{center}

Each RR~Lyr star has a unique identifier which follows the scheme
introduced in the OGLE-III catalogs. The identifiers from
OGLE-LMC-RRLYR-00001 to OGLE-LMC-RRLYR-24906 (in the LMC) and from
OGLE-SMC-RRLYR-0001 to OGLE-SMC-RRLYR-2475 (in the SMC) are reserved for
the RR~Lyr stars presented by Soszyñski \etal (2009, 2010). The identifiers
with higher numbers are assigned to the newly detected RR~Lyr stars in
order of increasing right ascension.

Our collection contains not only the most important parameters of the
sources (their coordinates, modes of pulsation, periods, mean magnitudes in
the {\it I}- and {\it V}-bands, amplitudes, and Fourier coefficients of the
light curve decomposition), but also the time-series {\it VI} photometry
collected from the beginning of the OGLE-IV survey. This photometry can be
combined with the OGLE-III and OGLE-II light curves (if available) from the
Soszyñski \etal (2009, 2010) catalogs, however in individual cases one
should compensate possible differences in the mean brightness and
amplitudes between the previous and present stages of the OGLE
project. About 8\% of the RR~Lyr stars included in the OGLE-III catalogs do
not have OGLE-IV photometry, mostly because they fell in technical gaps
between CCD chips of the OGLE-IV mosaic camera. The parameters of these
variables were copied from the OGLE-III catalog. In the future, we plan to
obtain the OGLE-IV time-series photometry also for the stars in the gaps,
since these regions are observed from time to time due to imperfections of
the telescope pointing.

\Section{Completeness of the Sample}
Due to a partial overlap of the adjacent OGLE-IV fields, some of the
sources were recorded twice, independently in both fields. However, our
collection contains only one entry per star -- in the case of the double
detections we usually chose the one with the larger number of observing
points in its light curve. These independent identifications of the same
RR~Lyr stars may be used to estimate the completeness of our sample.

We expect that the OGLE collection of RR~Lyr variables is practically
complete in the central regions of the LMC and SMC, that were monitored
since 2001 by the OGLE-III and OGLE-IV surveys. The outer regions are
affected by the gaps between CCD detectors of the mosaic camera, which
reduce the completeness by about 7\%. The efficiency of our search for
RR~Lyr stars in the area covered by the pixels may be judged on the basis
of the double detections. Outside the OGLE-III fields, 1284 variables
from our sample had two entries in the OGLE-IV database (assuming that both
light curves must have at least 100 points), so we had a chance to find
2568 counterparts. We independently identified 2480 of them, which
implies the completeness of about 96\%.

The highest completeness is expected for RRab stars, due to their
characteristic sawtooth-like light curves. Indeed, the same method applied
to the fundamental-mode pulsators gives the completeness well above
98\%. For RRc and RRd stars we obtained the completeness of about 92\%,
which reflects the difficulties to distinguish between the overtone
pulsators and close binary systems.

We compared the OGLE collection with the sample of 9722 RR~Lyr stars in
the LMC published by Alcock \etal (2001). Our sample does not include 131
of these objects, of which 40 are not present in the OGLE-IV database (most
of them lies in the gaps between the CCD chips). The remaining 91 stars
classified by MACHO as RR~Lyr stars either clearly belong to other types of
variable sources or are constant stars.

Kim \etal (2014) published a list of periodic variable star candidates
detected from the EROS-2 LMC photometric database. These objects were
classified using an automatic random forest algorithm. The list of
potential RR~Lyr stars contains 6607 sources not discovered during the
previous stages of the OGLE survey or the MACHO project. We cross-matched
an early version of our collection of RR~Lyr stars in the LMC with the
sample published by Kim \etal (2014) and we found that as many as 4408
object were missed in our list. For 3234 of these stars we found their
counterparts in the OGLE-IV database within 1~arcsec search radius. We
carefully analyzed the light curves of these objects and found that 149 of
them indeed are probable RR~Lyr stars. Most of these overlooked variables
turned out to be RRc stars with noisy, nearly sinusoidal light curves,
sometimes affected by a small number of points in their light curves. We
supplemented our collection with these newly identified RR~Lyr
variables. In turn, we do not confirm the Kim \etal (2014) classification
for the remaining 3085 sources. For the majority of these objects we had
no doubt that we deal with eclipsing binaries, $\delta$~Sct stars,
Cepheids, or simply just constant stars. There is also a number of sources
in this group for which the OGLE light curves are too noisy to
unambiguously categorize their type of variability.

\Section{Discussion}
The present version of the OGLE collection of RR~Lyr stars in the
Magellanic Clouds is larger and purer than any other catalog of these
pulsators detected in any other environment. Therefore, our sample is an
ideal tool to study RR~Lyr stars themselves, as well as the structure of
the Magellanic Clouds and their interactions with each other and our
Galaxy. Below we present a few possible applications of our collection,
however we are far from being exhaustive.
\begin{figure}[p]
\includegraphics[width=12.7cm]{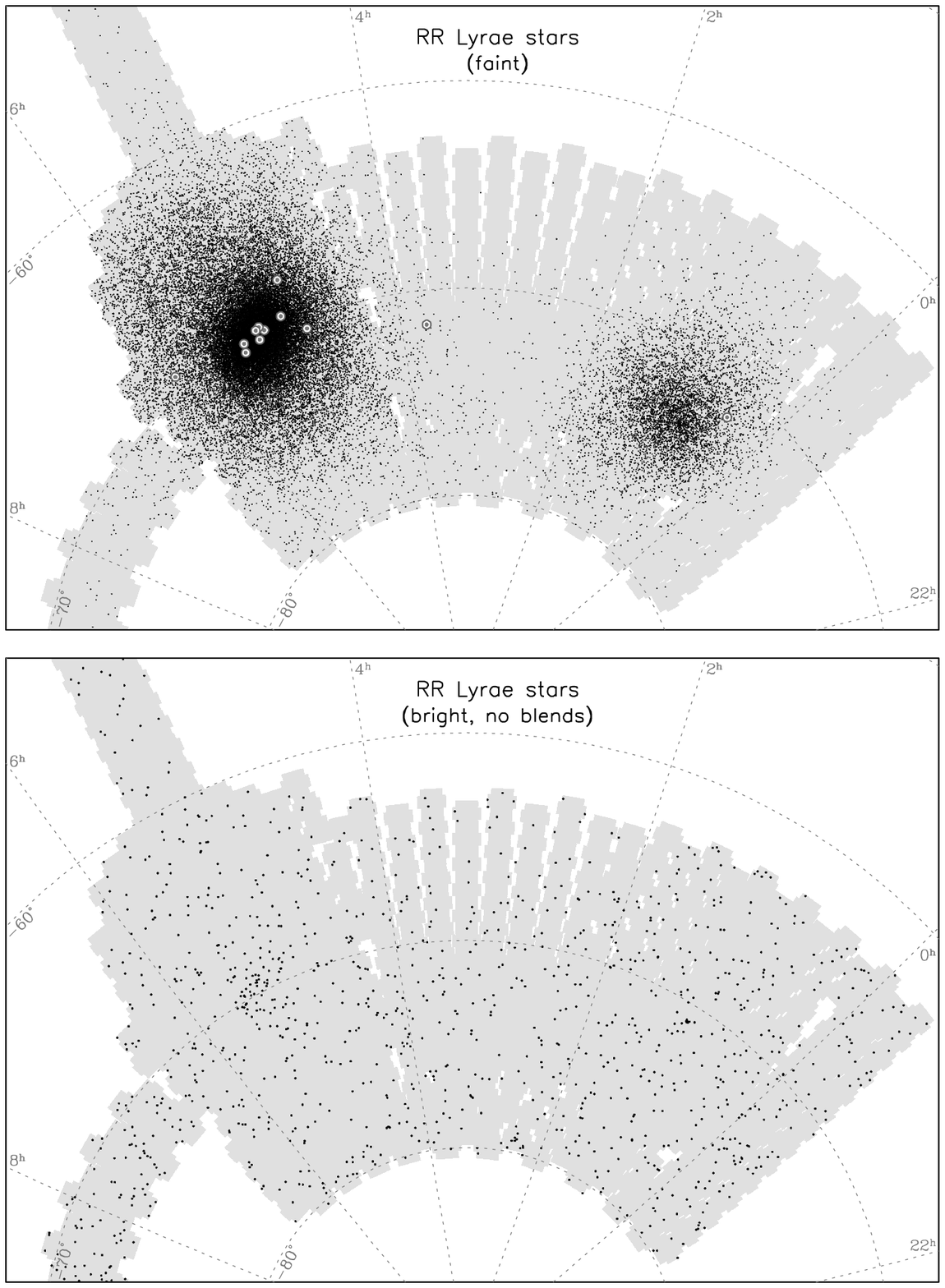}
\vspace*{4pt} 
\FigCap{Spatial distribution of RR~Lyr stars in the OGLE
  fields toward the Magellanic Clouds. {\it Upper panel} presents members
  of the Magellanic Clouds (fainter group). Gray circles indicate positions
  of twelve globular clusters that host RR~Lyr stars. {\it Lower panel} shows
  positions of the brighter group of RR~Lyr stars -- consisting mostly of
  the Milky Way members. The boundary between fainter and brighter groups
  has been adopted at 1~magnitude above the mean period--luminosity
  relations for RRab, RRc, and RRd stars, separately for the LMC and
  SMC. Additionally, blended stars have been removed from the bright
  group. Gray area shows the sky coverage of the OGLE fields.}
\end{figure}

\Subsection{Spatial Distribution of RR~Lyr Stars in the Magellanic System}
RR~Lyr stars are primary tracers of the ancient stellar population. These
pulsating stars are common in various environments, can be easy identified
in time-series sky surveys, and are standard candles, so can be used to
study the distribution of the old population in three dimensions. The
OGLE-III catalogs of RR~Lyr stars (Soszyñski \etal 2009, 2010) were
extensively used to investigate the structure of the Magellanic Clouds
(Pejcha and Stanek 2009, Subramaniam and Subramanian 2009, Haschke \etal
2012b, Kapakos and Hatzidimitriou 2012, Subramanian and Subramaniam 2012,
Wagner-Kaiser and Sarajedini 2013, Moretti \etal 2014, Deb and Singh 2014,
Deb \etal 2015). Now we extend the OGLE-III set to the peripheral areas of
the Clouds, which offers the opportunity to investigate the history of
interactions between the LMC, SMC, and Milky Way.

In Fig.~1, we present two-dimensional maps of RR~Lyr stars from our
collection. We divided our sample into fainter (upper panel of Fig.~1) and
brighter variables (lower panel), wherein the dividing lines were
arbitrarily defined at 1~magnitude above the mean period--luminosity
relations in the reddening-free Wesenheit index, $W_I=I-1.55(V-I)$,
separately for RRab, RRc, and RRd stars in the LMC and SMC. In the case of
the brighter group, we also cleaned the sample from blended variables. In
this procedure we relied on the light curve amplitudes, which are smaller
in blended variables than in typical pulsators with the same periods.

The fainter RR~Lyr stars belong mainly to the Magellanic Clouds, while the
brighter group is populated mostly by members of the Milky Way halo,
although there is no clear boundary between outer regions of these three
galaxies. This is seen in the distribution of the brighter RR~Lyr stars,
which is roughly uniform over the OGLE fields with the exception of the
center of the LMC. The excess of bright RR~Lyr stars in this region may be
partially explained by the contamination of not removed blends, but the
majority of these RR~Lyr stars have typical amplitudes of their light
curves, so they do not seem to be substantially blended. Thus, these stars
are likely located in the outskirts of the LMC stellar halo tidally
stretched toward our Galaxy.

The fainter group (upper panel of Fig.~1) mostly belong to the Magellanic
Clouds. The projection of the SMC halo on the celestial sphere seem to be
round, while the LMC halo is obviously elongated. Moreover, the
distribution of the LMC RR~Lyr stars probably cannot be described by a
simple ellipsoid, because the number of RR~Lyr stars in the North-East part
of the LMC seems to be larger than in the opposite side. A detailed
analysis of the three-dimensional distribution of RR~Lyr stars in the
Magellanic Clouds on the basis of our collection will be presented in the
forthcoming paper (Jacyszyn-Dobrzeniecka \etal in preparation).

\Subsection{RR~Lyr Stars in Globular Clusters}
The Oosterhoff dichotomy observed in Galactic globular clusters is not
present among the globular clusters in nearby dwarf galaxies, in particular
in the LMC. Five of the LMC clusters (NGC~1466, NGC~1853, NGC~2019,
NGC~2210, and NGC~2257) have properties that place them inside the zone of
avoidance between the two Oosterhoff groups in the Milky Way. This fact
poses a significant challenge to the models assuming hierarchical merger
formation of the Galactic halo.

The catalog of extended objects in the Magellanic System by Bica \etal
(2008) lists 18 globular clusters in the Magellanic Clouds, 14 of which are
currently monitored by OGLE. We found RR~Lyr stars in all but two of the
observed globular clusters. We found no RR~Lyr stars in Hodge~11 in the LMC
and Lindsay~1 in the SMC (strictly speaking we identified one RR~Lyr star
in Hodge 11, but it is probably a field variable located by chance in the
area outlined by the cluster radius). The simplest explanation for the lack
of RR~Lyr stars in Hodge~11 and Lindsay~1 is that these clusters are
younger than $\approx10$~Gyr.

\MakeTable{
l@{\hspace{8pt}} c@{\hspace{6pt}} c@{\hspace{6pt}} c@{\hspace{8pt}}
r@{\hspace{8pt}} r@{\hspace{6pt}}}{12.5cm}{Globular clusters containing
RR~Lyr stars}{\hline \noalign{\vskip3pt}
\multicolumn{1}{c}{Cluster name} & RA & Dec & Cluster & \multicolumn{1}{c}{$N_{\rm RR}$} & \multicolumn{1}{c}{$N_{\rm fieldRR}$} \\
  & (J2000) & (J2000) & radius [\arcm] & & (estimated) \\
\noalign{\vskip3pt}
\hline
\noalign{\vskip3pt}
NGC~121  &  00\uph26\upm47\ups  &  $-71\arcd32\arcm12\arcs$  &  3.8  &  15  &  1.2~~~~~ \\
NGC~1466 &  03\uph44\upm33\ups  &  $-71\arcd40\arcm17\arcs$  &  3.5  &  92  &  0.1~~~~~ \\
NGC~1754 &  04\uph54\upm17\ups  &  $-70\arcd26\arcm29\arcs$  &  1.6  &  37  &  0.4~~~~~ \\
NGC~1786 &  04\uph59\upm06\ups  &  $-67\arcd44\arcm42\arcs$  &  2.0  &  57  &  2.5~~~~~ \\
NGC~1835 &  05\uph05\upm06\ups  &  $-69\arcd24\arcm14\arcs$  &  2.3  & 125  &  6.4~~~~~ \\
NGC~1898 &  05\uph16\upm41\ups  &  $-69\arcd39\arcm23\arcs$  &  1.6  &  49  &  6.8~~~~~ \\
NGC~1916 &  05\uph18\upm38\ups  &  $-69\arcd24\arcm23\arcs$  &  2.1  &  25  & 10.5~~~~~ \\
NGC~1928 &  05\uph20\upm57\ups  &  $-69\arcd28\arcm40\arcs$  &  1.3  &   8  &  4.7~~~~~ \\
NGC~1939 &  05\uph21\upm26\ups  &  $-69\arcd56\arcm59\arcs$  &  1.4  &   7  &  4.9~~~~~ \\
NGC~2005 &  05\uph30\upm10\ups  &  $-69\arcd45\arcm10\arcs$  &  1.6  &  19  &  5.2~~~~~ \\
NGC~2019 &  05\uph31\upm56\ups  &  $-70\arcd09\arcm33\arcs$  &  1.5  &  61  &  4.6~~~~~ \\
NGC~2210 &  06\uph11\upm31\ups  &  $-69\arcd07\arcm18\arcs$  &  3.3  &  59  &  1.0~~~~~ \\
\noalign{\vskip3pt}
\hline}
Table~2 summarizes the properties of twelve globular clusters in the LMC and
SMC that host RR~Lyr stars. The coordinates and angular radii of the
clusters are taken from Bica \etal (2008). In the last two columns we
provide numbers of RR~Lyr stars detected within one radius from the
clusters' centers and estimated numbers of field RR~Lyr stars that are
expected to fall inside the same area. We estimated the number of field
variables counting RR~Lyr stars in the rings from 1.5 to 2.5 radii around
the cluster centers and rescaling these numbers to the area occupied by
clusters. The full lists of RR~Lyr stars found within the cluster radii are
provided in the FTP site in the file {\sf gc.dat}.

In two clusters -- NGC~1928 and NGC~1939 -- we found only eight and seven
RR~Lyr stars, respectively, while the expected number of field variables is
about five in both cases. Therefore, it cannot be excluded that all RR~Lyr
stars detected inside the area outlined by the radii of these clusters are
field variables. Spectroscopic and astrometric follow-up observations of
these stars should give a definitive answer to the question of their
membership. In other globular clusters listed in Table~2 the identification
of a significant number of RR~Lyr stars is firm, although in most cases
cluster members and field variables cannot be unambiguously distinguished.

\begin{figure}[tb]
\includegraphics[width=12.7cm]{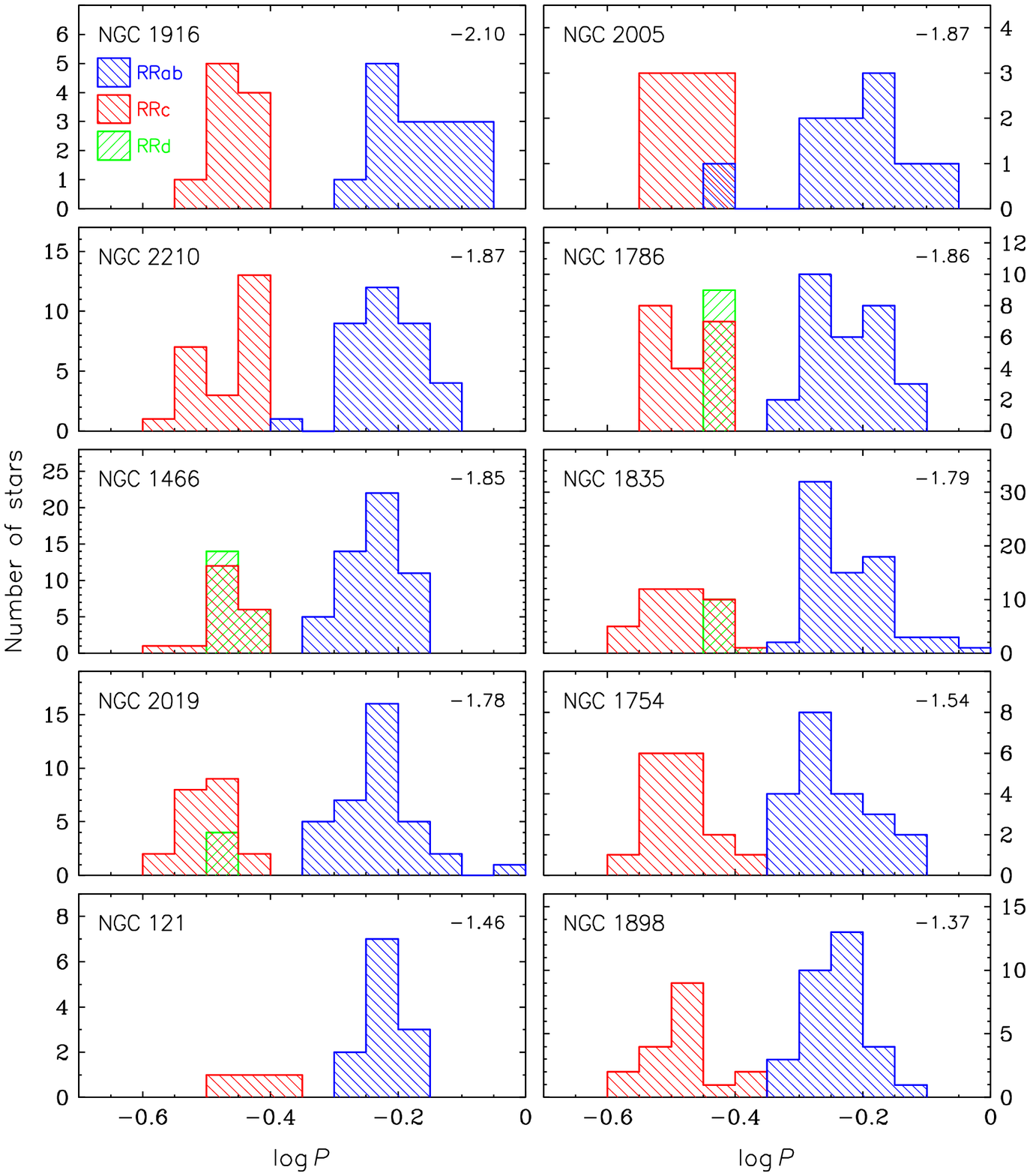}
\FigCap{Period distributions of RR~Lyr stars in the Magellanic Clouds'
  globular clusters. Blue, red, and green contours show histograms for
  RRab, RRc, and RRd (first-overtone periods) stars. Globular clusters are
  arranged by increasing metallicities, [Fe/H], given in the top right
  corner of {\it each panel}.}
\end{figure}
The distribution of pulsation periods of RR~Lyr stars in ten the richest
clusters are presented in Fig.~2. The clusters are arranged by increasing
metallicities to show the progression of the period distribution with
metallicity. RRd stars have been detected in four globular clusters and it
is interesting that all of them have intermediate metal abundances
$-1.86\le{\rm [Fe/H]}\le-1.78$ and fall in the Oosterhoff gap.

\Subsection{Double-Mode RR~Lyr Stars}
RR~Lyr stars with two first radial modes simultaneously excited (RRd stars)
constitute 5\% of the total sample in the LMC and 10\% in the SMC. These
are the largest sets of RRd stars known in any stellar environment, so they
may serve as important testbeds for theories of stellar pulsation. Petersen
diagram (period ratios plotted against the longer period) is a sensitive
tool widely used in asteroseismology (Popielski \etal 2000) and we
construct it to analyze RRd stars.

\begin{figure}[htb]
\includegraphics[width=12.5cm, bb=20 240 570 750]{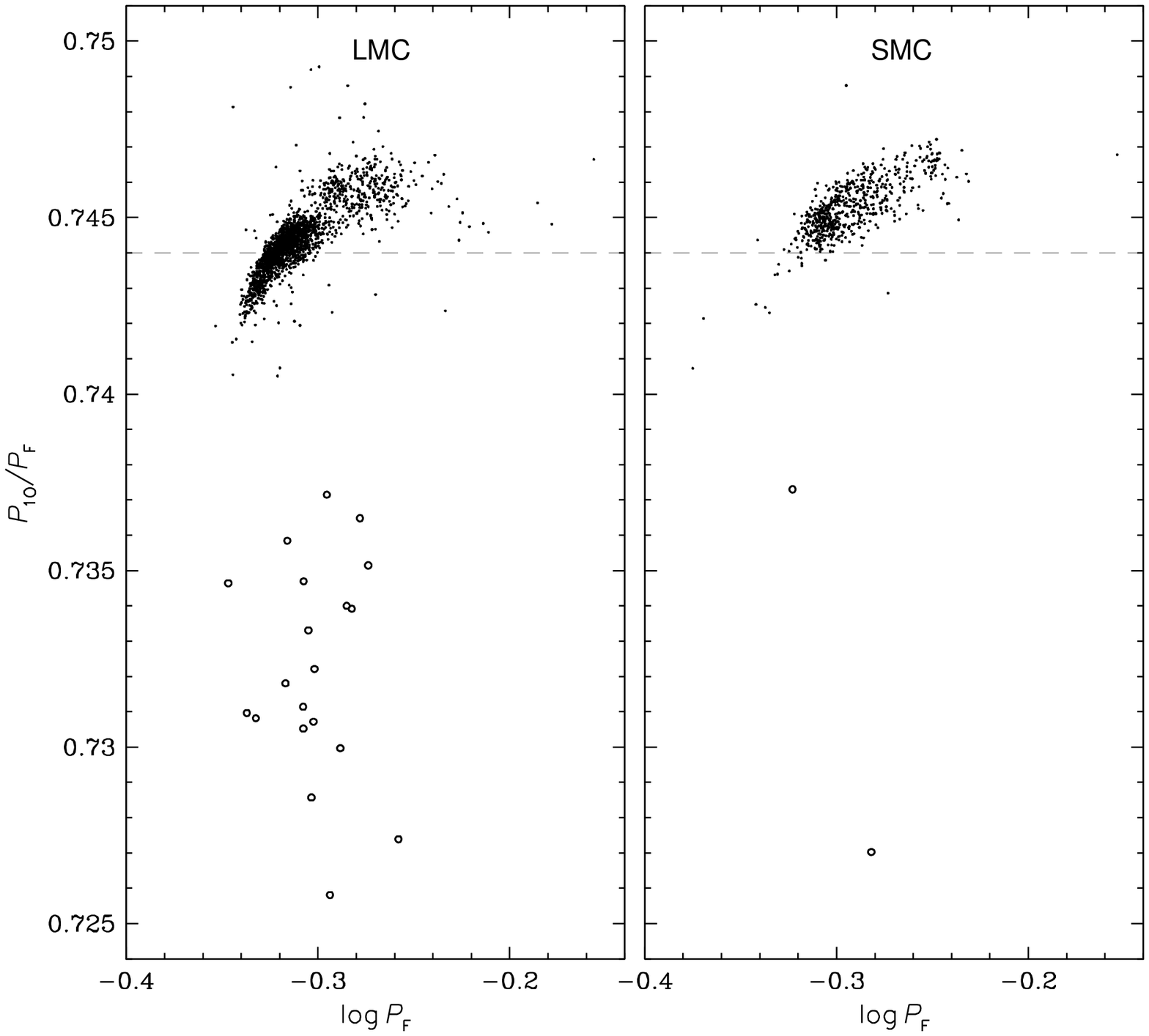}
\FigCap{Petersen diagrams for double-mode RR~Lyr variables in the LMC ({\it
    left panel}) and SMC ({\it right panel}). Black dots represent
  ``classical'' RRd stars, while empty circles mark anomalous RRd stars
  (Soszyñski \etal 2016, in preparation). Gray dashed lines indicate period
  ratio of 0.744 used to separate RRd stars shown in the two panels of
  Fig.~4.}
\end{figure}
\begin{figure}[p]
\includegraphics[width=12.7cm]{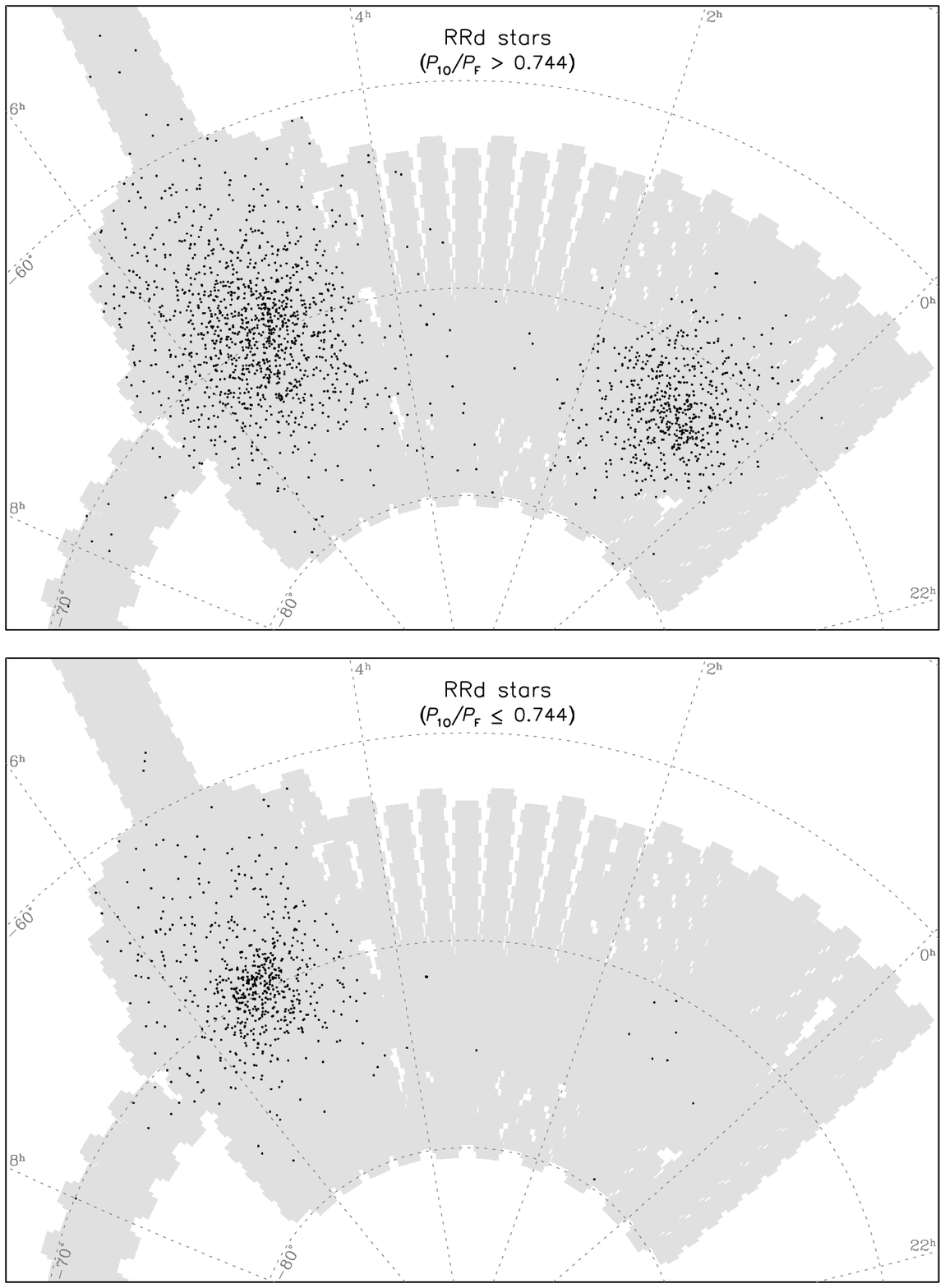}
\vspace*{4pt}
\FigCap{Spatial distribution of double-mode RR~Lyr stars in the OGLE fields
toward the Magellanic Clouds. {\it Upper panel} presents RRd stars with
period ratios $P_{\rm 1O}/P_{\rm F}>0.744$. {\it Lower panel} shows
positions of RRd stars with $P_{\rm 1O}/P_{\rm F}\le0.744$.}
\end{figure}
Fig.~3 shows the Petersen diagrams for RRd stars in the LMC and SMC. The
vast majority of double-mode RR~Lyr stars has period ratios within a narrow
range of $0.74<P_{\rm 1O}/P_{\rm F}<0.75$ and forms a curved sequence in
the diagram. The sequence is longer in the LMC and reaches smaller period
ratios than in the SMC. A simple test shows how the Petersen diagram is
sensitive to the chemical composition of the stars. We divided our sample
of RRd stars into two groups -- with period ratios above and below 0.744 --
and we checked the spatial distribution of both groups. The result is
displayed in Fig.~4. The $P_{\rm 1O}/P_{\rm F}\le0.744$ group is almost
absent in the SMC, while in the LMC both groups have clearly different
distributions, reflecting different metal abundance of these stars. RRd
stars with $P_{\rm 1O}/P_{\rm F}\le0.744$ are clearly more concentrated
toward the LMC center than the other group. This indicates that smaller
period ratios are associated with higher metal abundances.

In Fig.~3, we included a new class of double-mode RR~Lyr stars with period
ratios ranging between 0.725 and 0.738. We call these objects anomalous RRd
stars and describe them in the paper by Soszyñski \etal (2016, in
preparation). Anomalous RRd variables are characterized not only by
different ratios of periods in comparison to ``classical'' RRd stars, but
also by different amplitude ratios (in the anomalous RRd stars the
fundamental mode usually dominates) and different light curve morphology of
the fundamental mode component. Also, anomalous RRd stars usually show
modulations of the pulsation amplitudes, in other words -- the Blazhko
effect. First RRd stars exhibiting Blazhko modulation were recently
discovered in the Galactic bulge (Soszyñski \etal 2014, Smolec \etal 2015)
and these objects also belong to the anomalous subclass.

\Subsection{RR~Lyr Stars with Eclipsing Modulation}
In contrast to other types of classical pulsating stars (classical Cepheids
-- \eg\break Udalski \etal 2015b -- or type~II Cepheids), none of
RR~Lyr stars was confirmed as a member of an eclipsing binary system. In
the OGLE-II catalog, Soszyñski \etal (2003) discovered three RR~Lyr stars
in the LMC which show additional eclipsing variability superimposed on the
pulsation light curves. However, it is not clear whether the pulsating stars
are components of the binary systems, or these are physically unrelated
blends. Soszyñski \etal (2009) found one more RR~Lyr star with eclipsing
modulation. Very similar object -- OGLE-BLG-RRLYR-02792 -- detected by
Soszyñski \etal (2011) in the Galactic bulge was spectroscopically studied
by the Araucaria project (Pietrzyñski \etal 2012). They confirmed that
OGLE-BLG-RRLYR-02792 is indeed a pulsating star in a binary system, but it
cannot be a classical RR~Lyr star, since its mass is only 0.26~\MS. It
turned out that OGLE-BLG-RRLYR-02792 is a prototype of a new class of
internal variables -- binary evolution pulsators -- that mimic properties
and behavior of RR~Lyr stars.

\begin{figure}[p]
\centerline{\includegraphics[width=10.5cm]{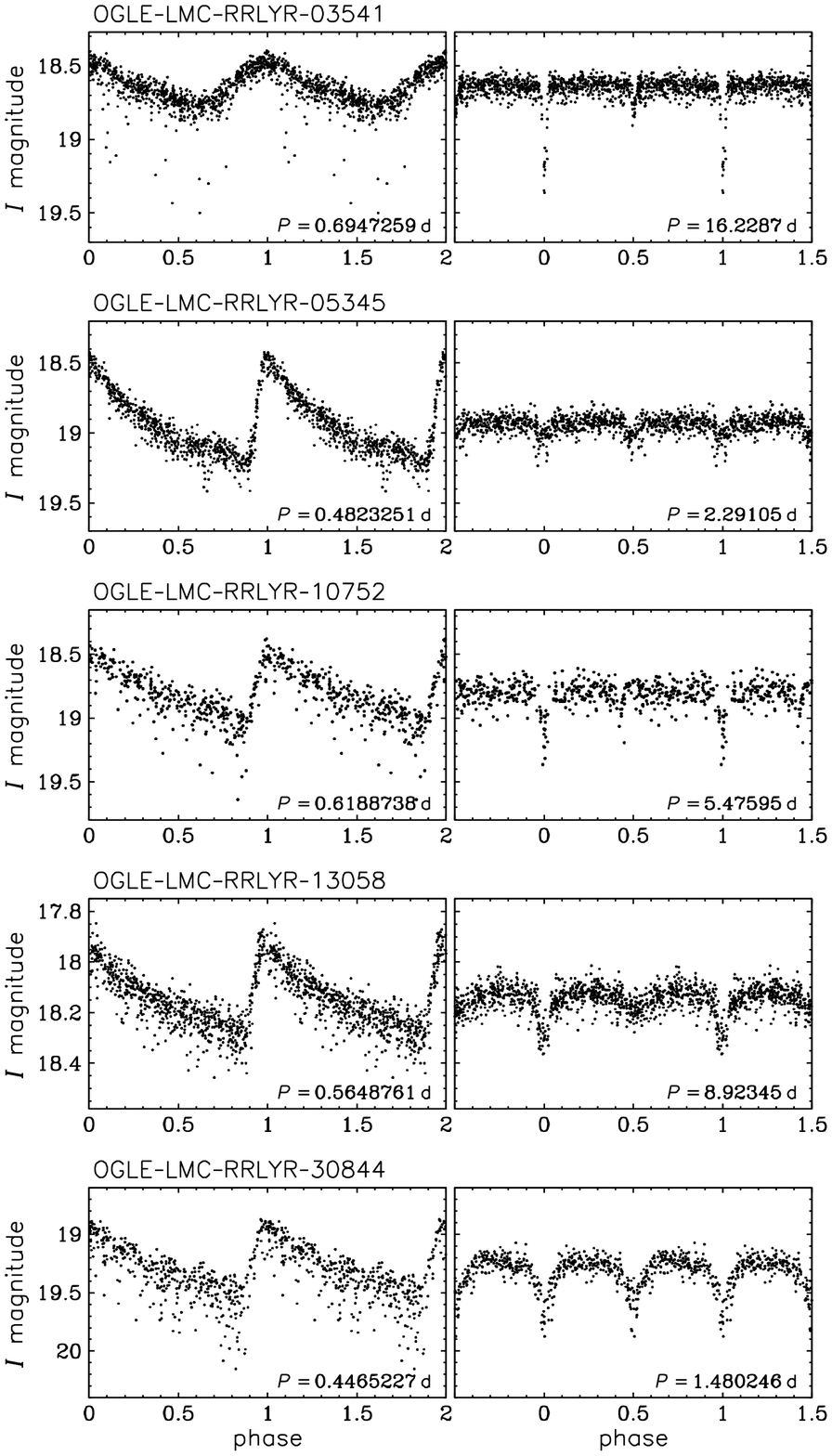}}
\vskip5pt
\FigCap{OGLE-IV {\it I}-band light curves of RR~Lyr stars showing eclipsing
variability. {\it Left panels} show the original photometric data folded
with the pulsation periods. {\it Right panels} show the eclipsing light
curves after subtracting the RR~Lyr component. The ranges of magnitudes are
the same in each pair of the panels.}
\end{figure}
In the present investigation, we report the discovery of one more candidate
for an RR~Lyr star with eclipsing-like modulation -- OGLE-LMC-RRLYR-30844
-- and we confirm the four previously announced objects of this kind
(Soszyñski \etal 2003, 2009). All these stars are located in the
LMC. Fig.~5 shows the OGLE-IV {\it I}-band light curves of all five stars.

Based solely on the photometric data it cannot be judged whether these
objects are real binary systems with a pulsating star as one of the
components or the eclipsing binaries are not related to the RR~Lyr stars and
are just optical blends. It should be noted that the orbital periods of our
candidates are very short compared to the values expected for horizontal
branch stars which in the previous stage of their evolution were located at
the tip of the red giant branch. Recently, Hajdu \etal (2015) conducted a
search for binary RR~Lyr stars in the Galactic bulge using the OGLE
collection (Soszyñski \etal 2011, 2014) and found 12 firm candidates for
(non-eclipsing) binary systems with the RR~Lyr components. The orbital
periods of all these candidates are longer than 1000~d. Our eclipsing
binaries have orbital periods between 1.48~d and 16.23~d, which may suggest
that we deal with optical blends.

However, the case of OGLE-BLG-RRLYR-02792 ($P_{\rm orb}=15.24$~d) indicates
that at least some of the stars shown in Fig.~5 may be binary evolution
pulsators -- stars that transferred most of their mass to their companions
and currently they cross the pulsation instability strip in their fast
evolution toward the helium white dwarf branch. In particular,
OGLE-LMC-RRLYR-03541 ($P_{\rm orb}=16.23$~d) has very similar properties
(period, light curve shape) to OGLE-BLG-RRLYR-02792. In turn,
OGLE-LMC-RRLYR-10752 exhibits a monotonic decrease of the pulsation period,
just as it is expected for fast-evolving binary evolution pulsators
(Pietrzyñski \etal 2012). Using the 18-years-long OGLE light curve of
OGLE-LMC-RRLYR-10752 we found the period change rate equal to
$-0.09\pm0.01$~s/yr.
\vspace*{-5pt}
\Section{Conclusions}
\vspace*{-3pt}
We presented the OGLE collection of over 45\,000 RR~Lyr stars in the
Magellanic System. Our sample contains, in fact, the vast majority of all
RR~Lyr variables in the Magellanic Clouds. A comparison of the OGLE-IV set
to the previous editions of the OGLE collection of variable stars
(Soszyñski \etal 2009, 2010) assures us that our collection is
characterized by a very high level of completeness and low
contamination. Therefore, it is currently the best suited dataset for
studying old stellar population in the Magellanic System. Distance
determinations, three-dimensional distribution of the ancient stars,
history of interactions between the Magellanic Clouds and the Milky Way,
mapping of the interstellar extinction, metallicity gradients in the
galaxies, properties of globular clusters -- these are just the most
obvious applications of the OGLE collection of RR~Lyr stars.

The sample of RR~Lyr stars itself is also a gold mine for stellar
astrophysics. Exotic multimode pulsations, non-radial modes, period
changes, mode switching, Blazhko effect, searching for pulsating stars in
binary systems -- all these studies are possible with the long-term OGLE
photometry published with this collection.
\vspace*{-5pt}

\Acknow{We would like to thank Profs. M.~Kubiak and G.~Pietrzyñ\-ski,
  former members of the OGLE team, for their contribution to the collection
  of the OGLE photometric data over the past years. We are grateful to
  Z.~Ko³aczkowski and A.~Schwar\-zen\-berg-Czerny for providing software
  used in this study.

  This work has been supported by the Polish Ministry of Science and Higher
  Education through the program ``Ideas Plus'' award No.~IdP2012
  000162. The OGLE project has received funding from the Polish National
  Science Centre grant MAESTRO no. 2014/14/A/ST9/00121 to AU. We
  acknowledge financial support from the National Science Centre grant
  No. DEC-2011/03/B/ST9/02573.}

\end{document}